\documentclass[pra,preprint,showpacs]{revtex4}
\usepackage[centertags]{amsmath}
\usepackage{amsfonts}
\usepackage{amssymb}
\usepackage{amsthm} 
\usepackage{newlfont}
\usepackage{epsfig}
\usepackage{amscd}
\usepackage{graphicx}
\usepackage{epsfig}
\usepackage{footnote}
\usepackage{lipsum}
\usepackage{color}
\usepackage{xcolor}
\usepackage{graphicx}
\usepackage{subfig}
\usepackage{amscd}

\newcommand{\beq}{\begin{equation}}
\newcommand{\eeq}{\end{equation}}
\newcommand{\ba}{\begin{array}}
\newcommand{\ea}{\end{array}}
\newcommand{\bea}{\begin{eqnarray}}
\newcommand{\eea}{\end{eqnarray}}

\begin{document}

\begin{center}
{\large \sc \bf {Two-channel spin-chain communication line and simple quantum gates}
}

\vskip 15pt

{\large 
J.Stolze$^1$ and A.I.~Zenchuk$^2$ 
}

\vskip 8pt

{\it $^1$Technische Universit\"at Dortmund, Fakult\"at Physik, D-44221 Dortmund, Germany},\\
{\it $^2$Institute of Problems of Chemical Physics, RAS,
Chernogolovka, Moscow reg., 142432, Russia}.

\end{center}


\begin{abstract}
We consider the  remote creation of a mixed state in a one-qubit
receiver connected to two  
two-qubit senders via different channels. {Channels are assumed to
  be chains of spins (qubits) with nearest-neighbor interactions, no
  external fields are being applied.} The problem of sharing the
creatable region of the receiver's state-space  
between two senders is considered for a communication line with the
receiver located asymmetrically with respect to  these senders
(asymmetric communication line). An example of a  quantum
register  realizing simple functions is constructed on the basis of a
symmetric communication line. In that setup, the initial states of
the two senders serve as input and  control signals, respectively,
while the state of the receiver at a proper time instant is considered
as the output signal.  
\end{abstract}

\maketitle

\section{Introduction}
\label{Section:Introduction}

The  remote creation of mixed states including the  control of  their
state parameters has become an attractive area of quantum information
processing.  
Originating from quantum state transfer (first explored by Bose
\cite{Bose}) this field of research aims at extending the possible
applications of quantum communication lines, stressing the quantum
nature of  the systems considered whose states are (generally) mixed
and thus described by density matrices. The power of the density
matrix lies in its large dimensionality even for  systems of small
size (for instance, the density matrix of  $N$  spins 1/2 has
dimension $2^N \times 2^N$) and, in addition, in the intricate
connection between the parameters  of the density matrix (changing one
of the parameters, in general, causes  changes in all others), which
is reflected in such characteristics of quantum systems as quantum
entanglement \cite{Wootters,HW} and discord \cite{HV,OZ,Z}. Therefore,
quantum operations based on the  density matrix of a mixed state seem
to offer a promising perspective in the development of quantum information devices.    

The research on  remote state creation started with photon systems
\cite{PBGWK2,PBGWK,XLYG}. Other two-level quantum systems were studied
less intensively (perhaps because of the obstacles in experimental
realization), although there has been some progress in that direction \cite{LH}.  In this regard we mention our recent papers on  one-qubit \cite{Z_2014,BZ_2015,BZ_2016} and two-qubit \cite{SZ_2016} state creation.

In order to optimize remote  state-creation protocols we use a  boundary controlled spin chain for state
creation. Originally it was observed that quantum state transfer can be
improved if the two sites at opposite ends of a chain are only weakly
coupled to the intermediate section \cite{GKMT,WLKGGB}. Later
performance was improved via optimizing one pair
\cite{BACVV,ZO,BACV,SAOZ}  or even two pairs of boundary coupling
constants \cite{ABCVV,SZ_2016}. In the present paper we will only
employ the second approach, that is, two-pair boundary-controlled chains.

{Since many of the references cited above deal with perfect state
transfer (PST) we would like to stress that we are presently considering remote
state creation. These two processes differ in various respects. In PST
a pure quantum state of (usually) a single qubit is identically
transferred to a different qubit. Remote state creation in contrast
deals with mixed states of sender and receiver units which may consist
of one or more qubits. In order to create a desired state of the
receiver, the sender first has to be prepared in an appropriate
initial state. Subsequently the desired state of the receiver develops
by the natural dynamics of the system, without external
control. Sender and receiver units (which need not be of the same
size) are connected by chains of spins 1/2 with nearest-neighbor
coupling. In contrast to PST the initial state of the sender and the
final state of the receiver usually are neither identical nor even
similar to each other.}

We pursue the idea of sharing the receiver's state space between two
different senders of a communication line. 
If the subregions creatable by different senders do not overlap, then
the created state can serve, in particular, as an indicator telling us
which of the two senders is involved in this process of state
creation. In our model, each of the  two-qubit senders is connected to
the one-qubit receiver by a  channel of spin-1/2 particles. Each
channel is an optimized boundary controlled chain with two pairs of
properly adjusted boundary coupling constants. This optimization
reduces significantly  the dependence of the creatable region on the
length of the channel. To demonstrate this, two communication lines with 
20 and 60 node channels are compared. Non-overlapping sharing of the
creatable  region requires an asymmetric communication line, for
instance, with different lengths of the two channels involved. 
Using, by contrast,  a similar  communication line with two equivalent
channels we  construct 
two quantum gates performing simple  manipulations with the eigenvalues
and eigenvectors of the receiver's density matrix.  

The paper is organized as follows. Two general schemes of  a
communication line together with the Hamiltonian governing the spin dynamics are discussed in Sec.\ref{Section:general}. The general settings for the remote state creation in  the two-channel communication line are given in Sec.\ref{Section:generalProtocol}. The sharing of the creatable region between two senders is studied in Sec.\ref{Section:nonsymm}.
Simple quantum gates are constructed in Sec.\ref{Section:gates} while
Sec.\ref{Section:conclusion} contains a general discussion.

\section{Two general schemes for a  two-channel communication line} 
\label{Section:general}
\begin{figure*}
\hspace*{-0.3cm}  \epsfig{file=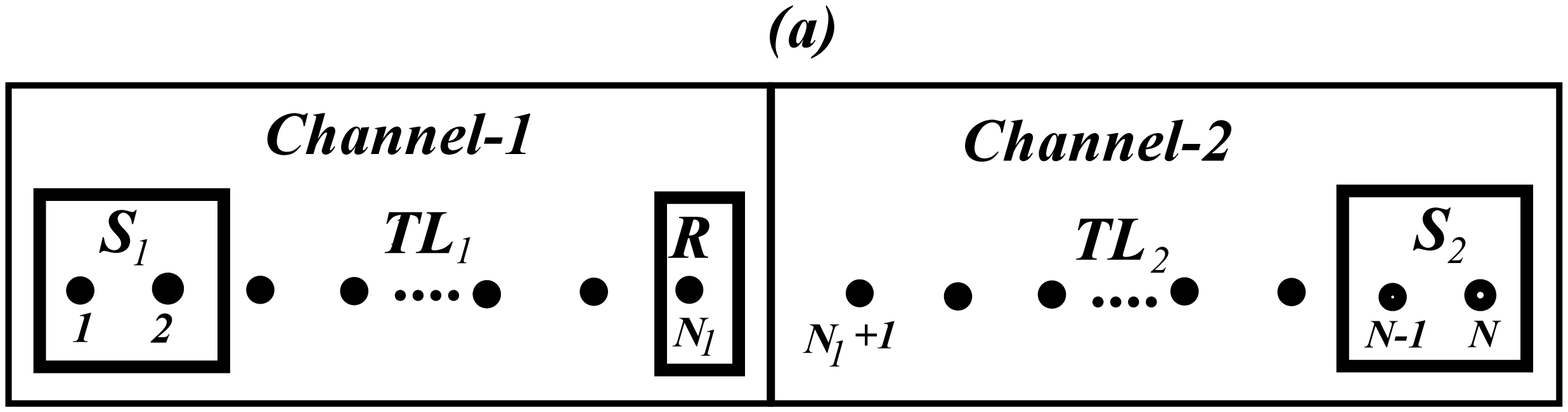, 
  scale=0.5
   ,angle=0
}  \\
\hspace*{-0.3cm}  \epsfig{file=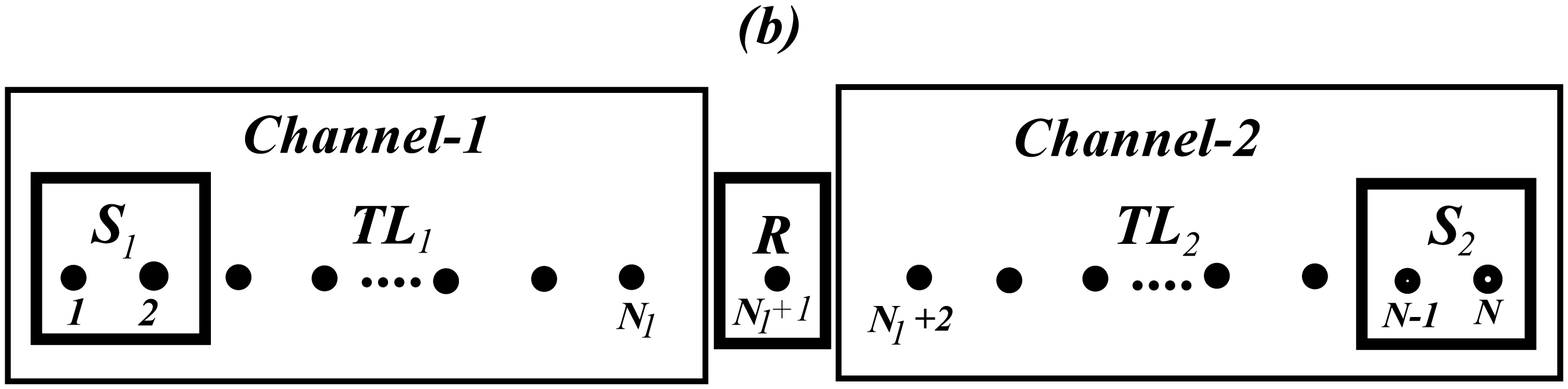, 
  scale=0.5
   ,angle=0
}  
\caption{The two-channel communication lines with the two-node senders
  $S_1$ and $S_2$ and the one-node receiver $R$ (a) embedded in the
  first channel (the communication line consists of $N=2 N_1$ nodes in
  total) and (b) located between the two channels (the communication
  line consists of $N=2 N_1+1$ nodes in total)}
  \label{Fig:2channel} \label{Fig:2channel2}
\end{figure*}

 We consider  two schemes for a two-channel communication line:
(i) the asymmetric scheme with the   receiver  at the last node of the first channel, 
Fig.\ref{Fig:2channel2}a, and (ii)  the symmetric scheme with the
receiver inserted between two equivalent channels, Fig.\ref{Fig:2channel}b.

If we want to separate the regions creatable by each of the senders
(while the other sender is in its ground state initially) 
then the asymmetric scheme (Fig.\ref{Fig:2channel}a) is preferable. By
contrast, the symmetric scheme will be useful for the quantum gates to
be discussed in Sec.\ref{Section:gates}. 

\subsection{Asymmetric communication line}
\label{Section:nonsymmetrical}
\begin{figure*}
\hspace*{-0.3cm}  \epsfig{file=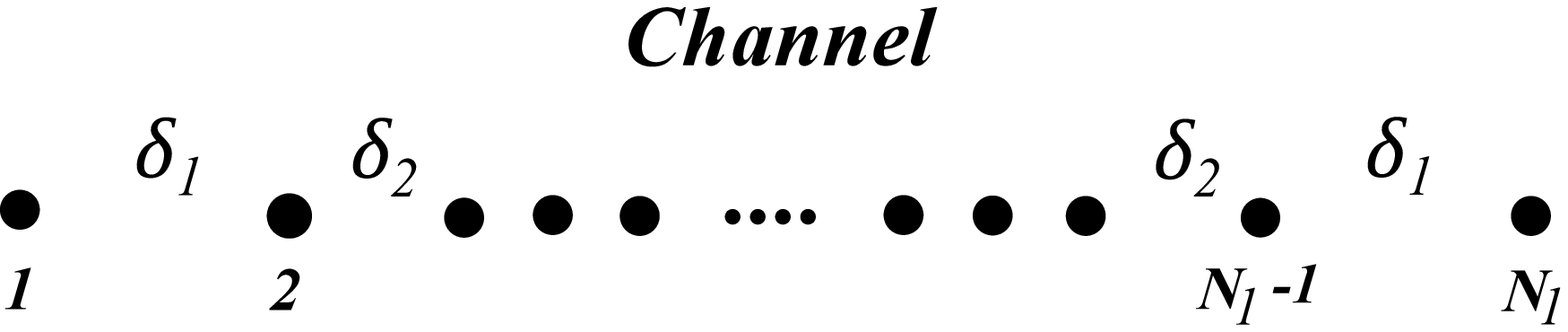, 
  scale=0.5
   ,angle=0
}  
\caption{The channel used as a building block in the  communication lines in Figs.\ref{Fig:2channel}(a) and \ref{Fig:2channel2}(b). There are two pairs of adjusted coupling constants: $D_1=D_{N_1-1}=\delta_1$ and 
$D_2=D_{N_1-2}=\delta_2$. All other coupling constants are equal,  $D_i=1$,  $2<i<N_1-2$}
  \label{Fig:channel} 
\end{figure*}

The communication line in Fig.\ref{Fig:2channel}(a) consists of two
equivalent channels, each with two pairs of properly adjusted end-nodes 
 \cite{SZ_2016} as shown in Fig.\ref{Fig:channel}.  
The whole system is governed by a nearest-neighbor XY-Hamiltonian with three parts:
\begin{eqnarray}\label{XY}
H=H_1+H_2+H_{12},
\end{eqnarray}
where $H_i$, $i=1,2$, are the Hamiltonians governing the dynamics within each particular channel and $H_{12}$ is the Hamiltonian responsible for the channel interaction:
\begin{eqnarray}\label{XY123}
H_1&=&\sum_{i=3}^{N_1-3} D (I_{ix}I_{(i+1)x}+I_{iy}I_{(i+1)y})+
\delta_1 \Big(I_{1x}I_{2x}+I_{1y}I_{2y} + I_{(N_1-1)x}I_{N_1x}+I_{(N_1-1)y}I_{N_1y}\Big)
+\\\nonumber
&&
\delta_2 \Big(I_{2x}I_{3x}+I_{2y}I_{3y} + I_{(N_1-2)x}I_{(N_1-1)x}+I_{(N_1-2)y}I_{(N_1-1)y} \Big),
      \\\nonumber
H_2&=&
\sum_{i=N_1+3}^{2 N_1-3} D (I_{ix}I_{(i+1)x}+I_{iy}I_{(i+1)y}) + \\\nonumber
&&
\delta_1 \Big(
          I_{(N_1+1)x}I_{(N_1+2)x}+I_{(N_1+1)y}I_{(N_1+2)y} + I_{(2 N_1-1)x}I_{(2 N_1)x}+I_{( 2 N_1-1)y}I_{(2 N_1)y}\Big)
+\\\nonumber
&&
\delta_2 \Big(
          I_{(N_1+2)x}I_{(N_1+3)x}+I_{(N_1+2)y}I_{(N_1+3)y} + I_{(2 N_1-2)x}I_{(2 N_1-1)x}+I_{( 2 N_1-2)y}I_{(2 N_1-1)y}\Big),\\\nonumber
H_{12}&=& \delta_3 \Big(I_{N_1x}I_{(N_1+1)x}+I_{N_1y}I_{(N_1+1)y}\Big)
\end{eqnarray}
The system thus consists of two subsystems of $N_1$ sites each. Note
that the subsystems are completely identical physically; the asymmetry
solely consists in identifying the end node of the left chain (site $N_1$) as
the receiver $R$ which receives information from either of the two
two-qubit senders $S_1$ and $S_2$, while the other sender is assumed
to be in its ground state initially. Note further that each of the two
identical subsystems is again physically symmetric in itself, with couplings
$\delta_1$ at the first and last nearest-neighbor bonds, and
$\delta_2$ at the second and second to last bonds. The
remaining nearest-neighbor couplings within each subsystem are all
equal to $D$.
Henceforth we consider the dimensionless time $D t$ formally assuming
$D=1$ in eq.(\ref{XY}). The two channels in this scheme are connected via the interaction between the $N_1$th and $(N_1+1)$th
nodes with the coupling constant $\delta_3$.

\begin{table}
\begin{tabular}{|ccccc|}
\hline
 $N_1$             &$\delta_1$ & $\delta_2$&$\delta_3$ &$t_0$ \cr 
\hline
 20            & 0.550& 0.817&0.28&28\cr
 60           & 0.414& 0.720&0.20&72.45\cr
\hline
\end{tabular}
\caption{The coupling constants $\delta_i$, $i=1,2,3$, and the time instant $t_0$  optimizing the  state 
creation in the asymmetric communication line (shown in Fig. \ref{Fig:2channel}(a))
with  pairs of channels of $N_1=20$ and $60$ nodes, respectively.  
} 
   \label{Table1}
   \end{table}

   The parameters $\delta_i$, $i=1,2$, are such that they maximize the
   probability of the excited state transfer between the end nodes of
   each individual $N_1$-site channel; the optimal values for  channels of lengths $N_1=20$ and $60$ were found in ref.\cite{SZ_2016}. Then, 
   having fixed $\delta_1$ and $\delta_2$, we adjust the coupling constant $\delta_3$ and the time instant of the state registration $t_0$ 
   which provide separation of the subregions of the receiver's
   state-space  creatable by the two different senders while keeping the area of
   each of the  creatable subregions large, as shown in Sec.\ref{Section:nonsymm}, Figs.\ref{Fig:reg}a, \ref{Fig:reg2}a. 
   All the parameters $\delta_i$, $i=1,2,3$ and $t_0$ are collected in Table \ref{Table1}.

\subsection{Symmetric communication line}
\label{Section:symmetric}
The communication line in Fig. \ref{Fig:2channel}(b) again employs two
chains of $N_1$ sites each as building blocks, but now the one-site
receiver $R$ is situated between the two chains, so that the complete
system now has $2N_1+1$ sites. The Hamiltonian  has the same structure
(\ref{XY})  as before, with the same $H_1$, physically different 
$H_{12}$  due to the presence of the additional site $R$, and formally
different   $H_2$ due to a renumbering of the sites:

\begin{eqnarray}\label{XY123G}
H_2&=&
\sum_{i=N_1+4}^{2 N_1-2} (I_{ix}I_{(i+1)x}+I_{iy}I_{(i+1)y}) + \\\nonumber
&&
\delta_1 \Big(
          I_{(N_1+2)x}I_{(N_1+3)x}+I_{(N_1+2)y}I_{(N_1+3)y} + I_{(2 N_1)x}I_{(2 N_1+1)x}+I_{( 2 N_1)y}I_{(2 N_1+1)y}\Big)
+\\\nonumber
&&
\delta_2 \Big(
          I_{(N_1+3)x}I_{(N_1+4)x}+I_{(N_1+3)y}I_{(N_1+4)y} + I_{(2 N_1-1)x}I_{(2 N_1)x}+I_{( 2 N_1-1)y}I_{(2 N_1)y}\Big),\\\nonumber
H_{12}&=& \delta_3 \Big(I_{N_1x}I_{(N_1+1)x}+I_{N_1y}I_{(N_1+1)y}+I_{(N_1+1)x}I_{(N_1+2)x}+I_{(N_1+1)y}I_{(N_1+2)y}\Big).
\end{eqnarray}
The bulk coupling constant is again $D\equiv 1$ and the parameters
$\delta_1$ and $\delta_2$  are the same as before, see Table \ref{Table1},
 while the parameters   $\delta_3=0.318$ and  $t_0=29.190$ maximize
 the probability of the excited state transfer from  one of the
 senders to the receiver (different from  the parameters $\delta_3$ and $t_0$ in the asymmetric communication line, 
 Sec.\ref{Section:nonsymmetrical}, which were optimized according to
 different criteria).
 
 In
addition to the two different systems described above we introduce a
special symmetric system whose properties will be 
discussed in Sec. \ref{Section:gates} below. In that system the
two-node senders $S_1$ and $S_2$ are directly coupled to the receiver
$R$ without any intermediate channel. This minimal system then has
length $N=5$.
The corresponding nearest-neighbor XY Hamiltonian reads:
\begin{eqnarray}
\label{5sites}
H=   \Big(I_{1x}I_{2x}+I_{1y}I_{2y}+I_{4x}I_{5x}+I_{4y}I_{5y}\Big)+ \delta_3 \Big(I_{2x}I_{3x}+I_{2y}I_{3y}+
I_{3x}I_{4x}+I_{3y}I_{4y}\Big)
\end{eqnarray}
For this case,  the optimal parameters read:  $t_0=4.443$, $\delta_3=0.707$.

\section{General protocol of remote one-qubit state creation}
\label{Section:generalProtocol}

We now explain how to use the quantum hardware described in the
previous section.
Let us  consider the asymmetric  two-sender communication line with
two $N_1$-site channels shown in Fig.\ref{Fig:2channel2}a. 
Our protocol is based on the tensor-product  initial state 
\begin{eqnarray}\label{Psi0}
\rho_0 = |\Psi_{S_10}\rangle\langle \Psi_{S_10}| \otimes |\Psi_{rest}\rangle\langle \Psi_{rest}| \otimes |\Psi_{S_20}\rangle\langle \Psi_{S_20}|,
\end{eqnarray}
where
$|\Psi_{S_10}\rangle$, $|\Psi_{S_20}\rangle$ and $|\Psi_{rest}\rangle$
are, respectively, the normalized initial states of the first sender
$S_1$, of the second sender $S_2$ and of the rest of the communication
line (i.e., the two transmision lines $TL_1$, $TL_2$ together with the
receiver $R$):  
\begin{eqnarray}\label{inS}
&&
 |\Psi_{S_10}\rangle = a_{10}|0\rangle + a_{11} |1\rangle+ a_{12} |2\rangle,\\\label{inR}
 &&
|\Psi_{S_20}\rangle= a_{20}|0\rangle + a_{21} |N\rangle+ a_{22} |N-1\rangle,\\\label{inTL}
 &&
|\Psi_{rest}\rangle=|0\rangle.
\end{eqnarray}
Here $|0\rangle$ is the state without excitations and  $|k\rangle$
means the state with the $k$th spin excited. The coefficients in the
above states can be conveniently parametrized in terms of angles:
\begin{eqnarray}\label{inpar}
&&
a_{10}=\sin\frac{\alpha_{11} \pi}{2} ,\;\;\\\nonumber
&&
a_{11}= \cos\frac{\alpha_{11} \pi}{2} \cos\frac{\alpha_{12} \pi}{2} e^{2\pi i \phi_{11}} ,\\\nonumber
&&
a_{12}=\cos\frac{\alpha_{11} \pi}{2} \sin\frac{\alpha_{12} \pi}{2} e^{2\pi i \phi_{12}} ,\\\nonumber
&&
a_{20}=\sin\frac{\alpha_{21} \pi}{2} ,\;\;\\\nonumber
&&
a_{21}= \cos\frac{\alpha_{21} \pi}{2} \cos\frac{\alpha_{22} \pi}{2} e^{2\pi i \phi_{21}} ,\\\nonumber
&&
a_{22}=\cos\frac{\alpha_{21} \pi}{2} \sin\frac{\alpha_{22} \pi}{2} e^{2\pi i \phi_{22}},
\end{eqnarray}
with
\begin{eqnarray}\label{parameters}
0\le \alpha_{ij}\le 1,\;\;\;0\le \phi_{ij}\le 1,\;\;i,j=1,2.
\end{eqnarray}
Since the initial state involves at most two excited nodes and the XY Hamiltonian (\ref{XY123}) 
commutes with the $z$-projection of the total spin momentum $I_z$ ($[H,I_z]=0$) then  
the spin dynamics can be described in the subspace  spanned by the vectors 
\begin{eqnarray}\label{basis}
|0\rangle,\;\;|k\rangle, \;k=1,\dots,N, \; |ij\rangle,\; j>i,\;i,j=1,\dots,N
\end{eqnarray}
whose dimensionality is $N + 1 + \left({N}\atop{2}\right)= \frac{1}{2} (N^2 + N + 2)$.
Again, $|0\rangle$ is the state without excitations,  $|k\rangle$
means the state with the $k$th spin excited, and  
 $|ij\rangle $ is the state with the $i$th and
$j$th spins excited.

We represent the state of the receiver $R$ in the following factorized form:
\begin{eqnarray}\label{RhoR}
\rho^R=U\Lambda U^+,
\end{eqnarray}
where $\Lambda$ and $U$  are, respectively,  the diagonal matrix of the eigenvalues  of $\rho^R$ and  the matrix of its eigenvectors:
\begin{eqnarray}
\label{Lambda}
&&
\Lambda={\mbox{diag}} (\lambda,1-\lambda),
\\\label{U}
&&
U=\left(
\begin{array}{cc}
\cos \frac{\pi \beta_1}{2} & - e^{-2 i \pi \beta_2} \sin \frac{\pi \beta_1}{2}\cr
 e^{2 i \pi \beta_2} \sin \frac{\pi \beta_1}{2} & \cos \frac{\pi \beta_1}{2} 
\end{array}
\right) ,
\end{eqnarray}
and $\sigma_i$ are the Pauli matrices:
\begin{eqnarray}
\sigma_1=\left(
\begin{array}{cc}
0&1\cr
1&0
\end{array}
\right),\;\;
\sigma_3={\mbox{diag}}(1,-1).
\end{eqnarray}
Hereafter we study the creatable  region in the plane $(\lambda,\beta_1)$ of the receiver's state space disregarding the
value of the parameter $\beta_2$ for the sake of simplicity.

\section{Asymmetric communication line: sharing the creatable region between two senders}
\label{Section:nonsymm}

Every initial state (\ref{Psi0}) of the system is mapped to a final state (at time
$t_0$) (\ref{RhoR}) of the receiver, characterized by parameters
$\lambda$ and $\beta_1$. As the parameters $\alpha_{ij}$ and
$\phi_{ij}$ (\ref{inpar}) of the initial state vary continuously over
some region, the state of the receiver varies over a region in the
$(\lambda, \beta_1)$-plane, which we call the creatable region
corresponding to the given region of the $\alpha_{ij}$ and
$\phi_{ij}$. In this section we study the shape of the creatable
region for two types of initial states.

 \subsection{One-sided initial state}

\begin{figure*}
\subfloat[]{\includegraphics[scale=0.8,angle=0]{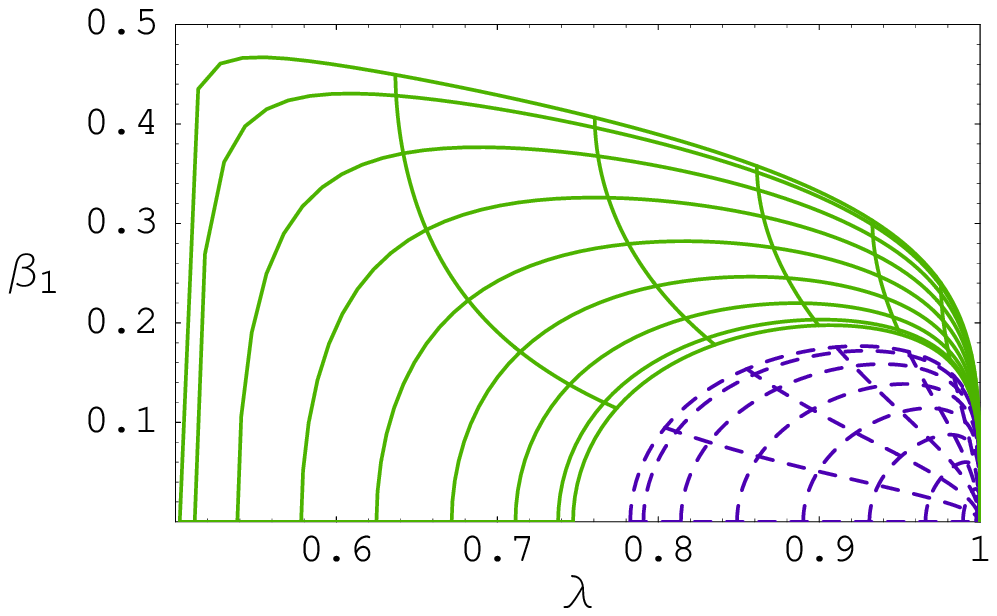}}
\subfloat[]{\includegraphics[scale=0.8,angle=0]{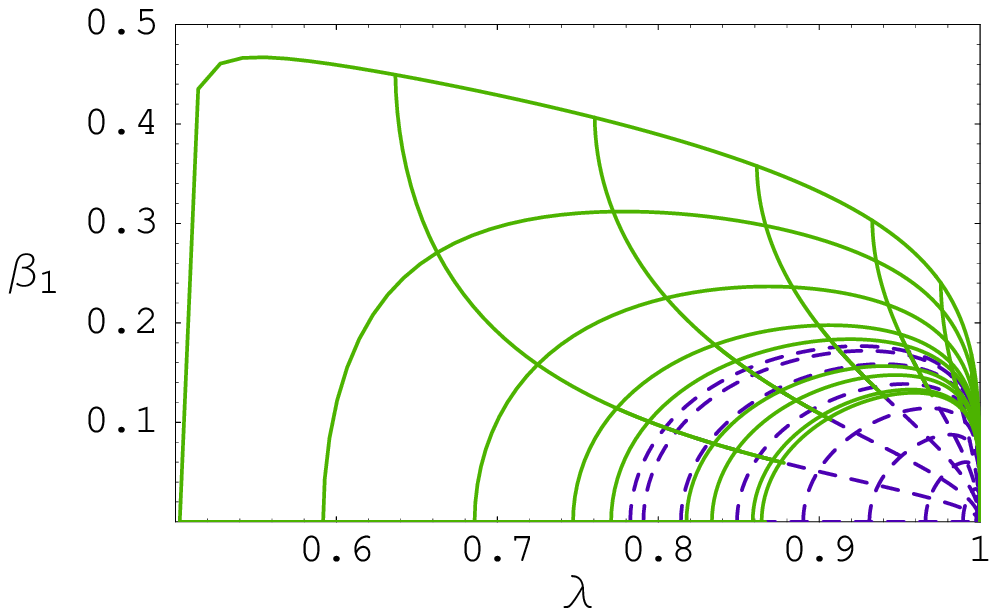}}\\
\subfloat[]{\includegraphics[scale=0.8,angle=0]{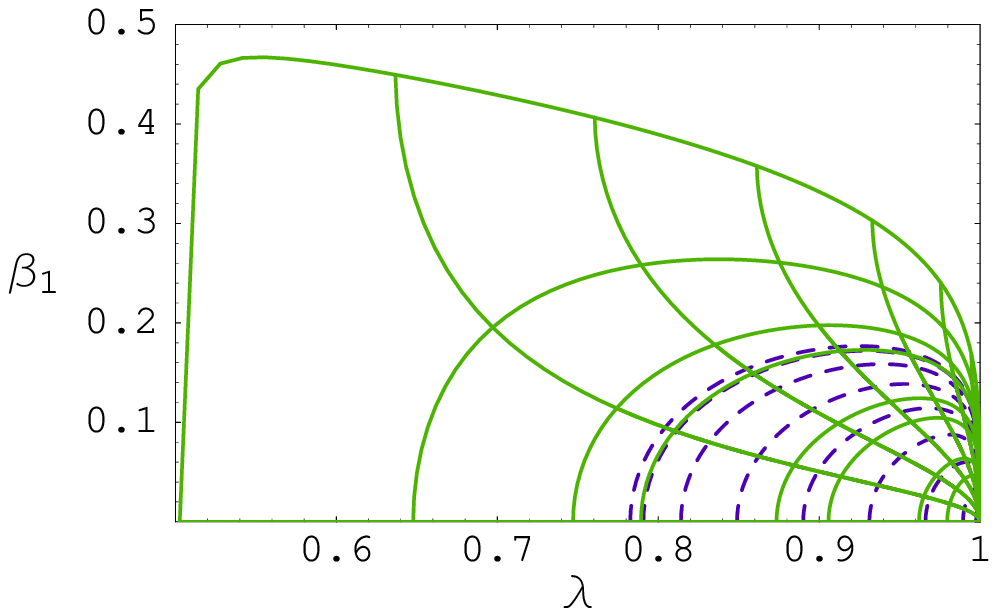}}
\caption{ The creatable regions of the receiver's state space in the
  plane $(\lambda, \beta_1)$ for the asymmetric communication line 
of $N=2 N_1=40$ nodes. 
The larger (solid lines)  and smaller (dashed lines) creatable regions correspond, respectively, to
$IS_{10}$ and $IS_{01}$ given in (\ref{IS1}). 
(a) $\phi_{12} = 0$,  the two  creatable regions do not
overlap. (b) $\phi_{12} = 0.6$, there is some overlap  of
the two creatable regions. (c) $\phi_{12} = {0.75}$, the smaller creatable region is
completely embedded in the larger one. }
  \label{Fig:reg} 
\end{figure*}

\begin{figure*}
\subfloat[]{\includegraphics[scale=0.8,angle=0]{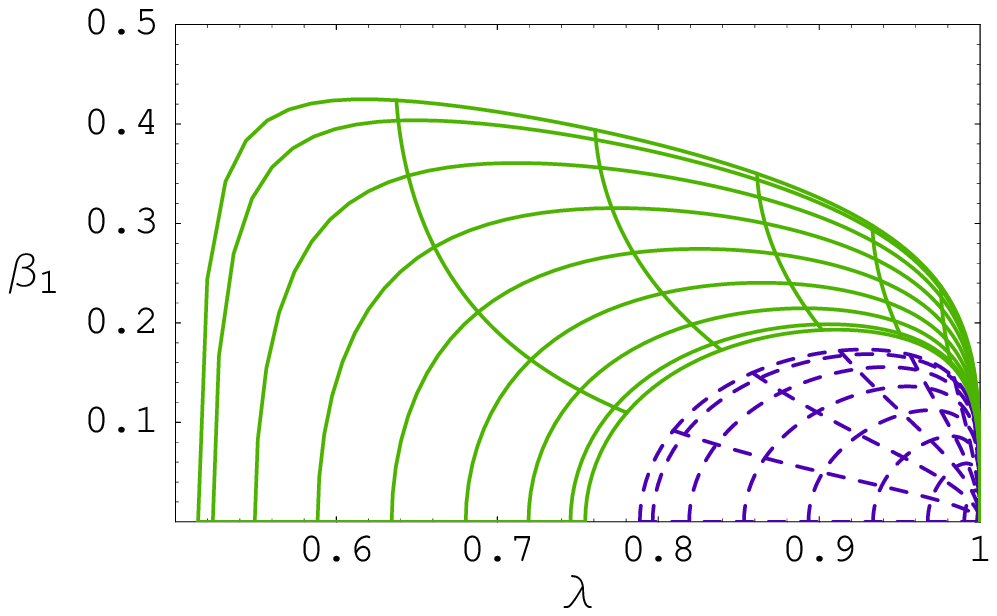}}
\subfloat[]{\includegraphics[scale=0.8,angle=0]{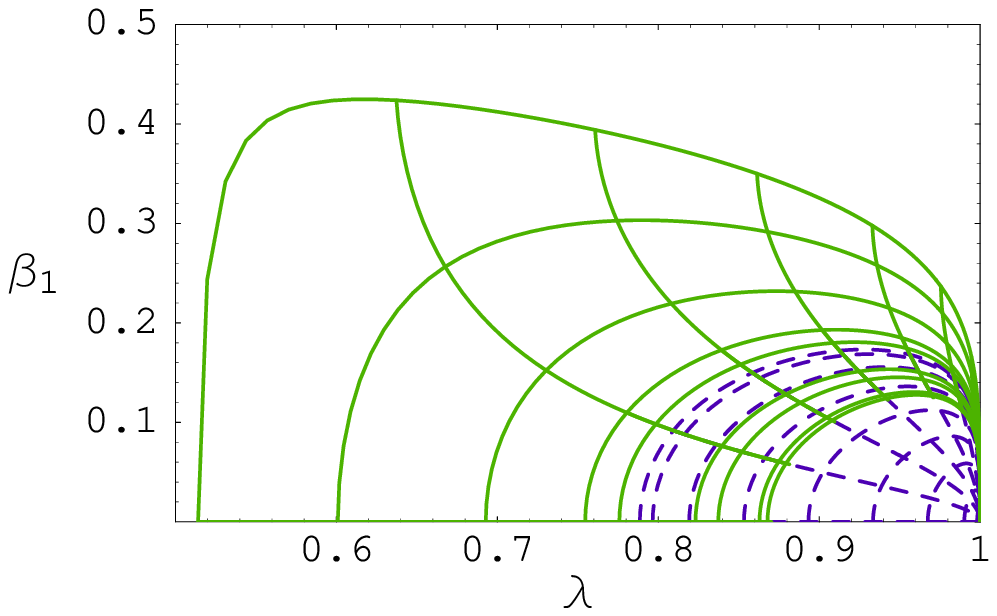}}\\
\subfloat[]{\includegraphics[scale=0.8,angle=0]{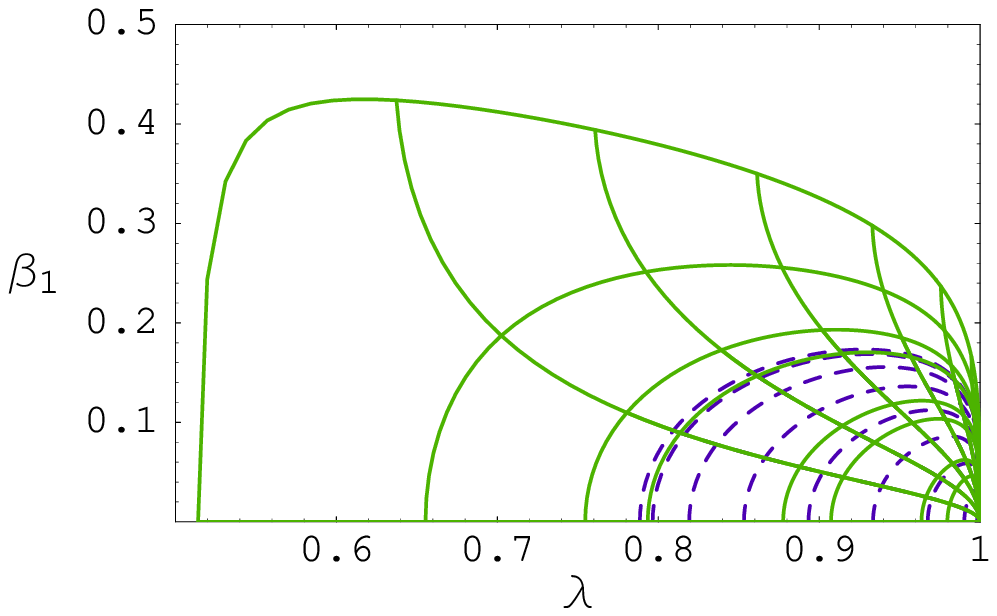}}
\caption{The same as in Fig.\ref{Fig:reg} 
for the communication line of 
$N = 2 N_1  = 120$ nodes. 
(a) $\phi_{12} = 0$, the two
creatable regions do not overlap. (b) $\phi_{12} = 0.6$, there is some overlap
  between the two creatable regions. (c)  $\phi_{12} = {0.75}$, the
smaller creatable region is completely embedded in the larger one.}
  \label{Fig:reg2} 
\end{figure*}

 First, we consider the two initial states , $IS_{10}$ and $IS_{01}$, where one of the senders is in the ground state:
 \begin{eqnarray}\label{IS1}
 &&
 IS_{10}: \;\;{\mbox{$|\Psi_{S_{2}0}\rangle= |0\rangle$ and  $|\Psi_{S_{1}0}\rangle$ is state (\ref{inS}) with 
 $\phi_{11}=0$} } \\\nonumber
 &&
 IS_{01}: \;\;{\mbox{$|\Psi_{S_{1}0}\rangle= |0\rangle$ and  $|\Psi_{S_{2}0}\rangle$ is state (\ref{inR}) with 
 $\phi_{2i}=0$, $i=1,2$} }
 \end{eqnarray}
 Thus, we keep $\phi_{12}$ in $IS_{10}$ free and use this parameter to
 control the shape of the creatable region, while the creatable region
 is fixed for $IS_{01}$. The creatable region is then mapped out by
 varying the parameters $\alpha_{11}$ and   $\alpha_{12}$  for state
 $IS_{10}$ and  $\alpha_{21}$ and   $\alpha_{22}$  for state
 $IS_{01}$, respectively. 
The
creatable regions corresponding to both of these initial states  are
shown in Fig.\ref{Fig:reg} ($N_1=20$) and Fig.\ref{Fig:reg2}
($N_1=60$). The structure of these regions can be read off from the
lines of constant  $\alpha_{11}$ and   $\alpha_{12}$ (solid lines) for state
 $IS_{10}$ and the lines of constant  $\alpha_{21}$ and
 $\alpha_{22}$ (dashed lines)  for state  $IS_{01}$. In particular,
 the { monotonic} solid and dashed lines emanating from the lower right corners of the figures
 correspond, respectively, to values $\alpha_{11}$ and $\alpha_{21}$ varying (from left to right) from 0 (these two
 lines coincide with the absciss axis)  to 1 with the interval $\Delta \alpha=\frac{1}{8}$. The remaining solid and dashed lines
 correspond to  values, respectively,  $\alpha_{12}$  and  $\alpha_{22}$ varying (from top to bottom) from 0 to 1 with the same 
 interval $\Delta \alpha$.
 Figs.\ref{Fig:reg}(a) and \ref{Fig:reg2}(a) show data for
 $\phi_{12}=0$. The two 
creatable subregions for $IS_{10}$ and $IS_{01}$ do not overlap so that
  the registered state informs us which sender was used for its
  creation. For $\phi_{12} = 0.6$,  Fig.\ref{Fig:reg}(b),
  \ref{Fig:reg2}(b), there is some overlap of the two creatable
  subregions so that the states in this overlap can be controlled by
  both senders. 
Finally, for $\phi_{12} = { 0.75}$ , Fig.\ref{Fig:reg}(c),
\ref{Fig:reg2}(c), the smaller creatable region is 
completely embedded in the larger one, so that all states in the
smaller subregion can be created by both senders.

\subsection{Two-sided initial state}

Now we consider initial states of the communication line in which
neither of the two
senders  $S_1$ and $S_2$ is in the ground state. In that situation one
of the senders can, by fixing its control parameters $\alpha_{ij},
\phi_{ij}$, be used to control the position and form of the creatable
region that the other sender can create as its initial state
parameters vary over some range.
As examples we consider the following three initial states:
\begin{eqnarray}\label{IS2}
IS_1: && \; {\mbox{$|\Psi_{S_10}\rangle$ is state (\ref{inS}) with $\phi_{1i}=0$ (input state);}}\\\nonumber
&& {\mbox{ $|\Psi_{S_20}\rangle$  is state (\ref{inR}) with $\alpha_{2i}=\phi_{2i}=0$ (controlling state)}},\\\nonumber
IS_2: && {\mbox{ $|\Psi_{S_10}\rangle$  is state (\ref{inS}) with $\alpha_{12}=\phi_{2i}=0$, $\alpha_{11}=\frac{1}{2}$ (controlling state)}},\\\nonumber
&& \; {\mbox{$|\Psi_{S_20}\rangle$ is state (\ref{inR}) with $\phi_{2i}=0$ (input  state);}}\\\nonumber
IS_3: && \; {\mbox{$|\Psi_{S_10}\rangle$ and $|\Psi_{S_20}\rangle$ are states, respectively,
(\ref{inS}) and (\ref{inR}) with $\alpha_{1i}=\alpha_{2i}$,  }}\\\nonumber && {\mbox{\hspace{3cm} ( {$\phi_{1i}=\phi_{2i}=0$}, equivalent initial states of $S_1$ and $S_2$)}}.
\end{eqnarray}
In $IS_1$ the sender $S_1$ is in a normalized superposition of
$|0\rangle$, $|1\rangle$, and $|2\rangle$ with arbitrary positive
coefficients (a convex combination), while the controlling state of
$S_2$ is fixed at $|N\rangle$. In $IS_2$, in contrast, $S_2$ varies
over all convex combinations of  $|0\rangle$, $|N-1\rangle$, and
$|N\rangle$ while $S_1$ is in the fixed state $ \frac
1{\sqrt{2}}\left(|0\rangle+|1\rangle\right)$.  Finally, in $IS_3$ $S_1$ and $S_2$ are assumed to be in
the same convex combination of $|0\rangle$, $|1\rangle$, and
$|2\rangle$ and $|0\rangle$, $|N\rangle$, and
$|N-1\rangle$, respectively. 

\begin{figure*}
\subfloat[]{\includegraphics[scale=0.8,angle=0]{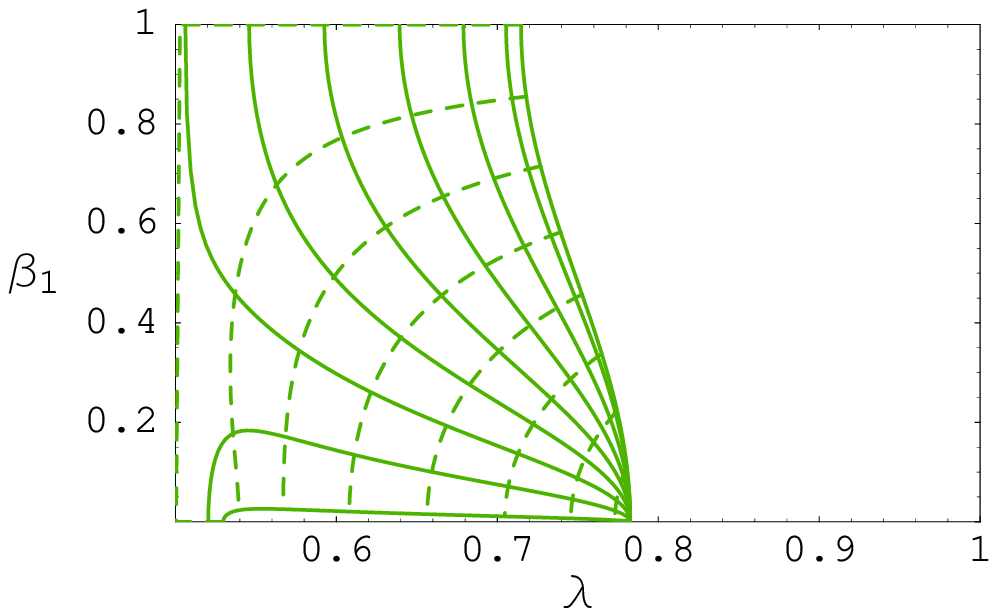}}
\subfloat[]{\includegraphics[scale=0.8,angle=0]{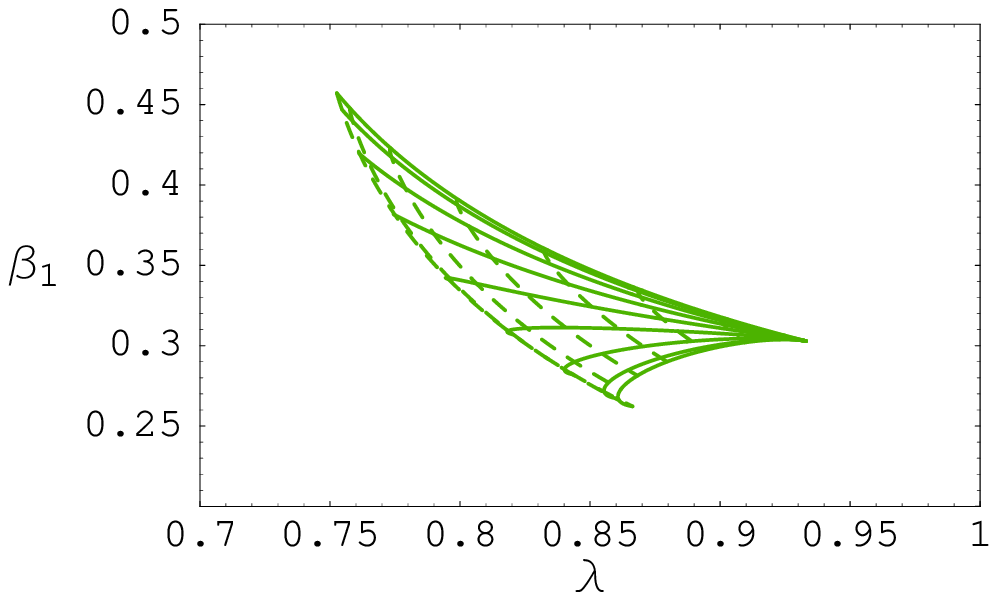}}\\
\subfloat[]{\includegraphics[scale=0.8,angle=0]{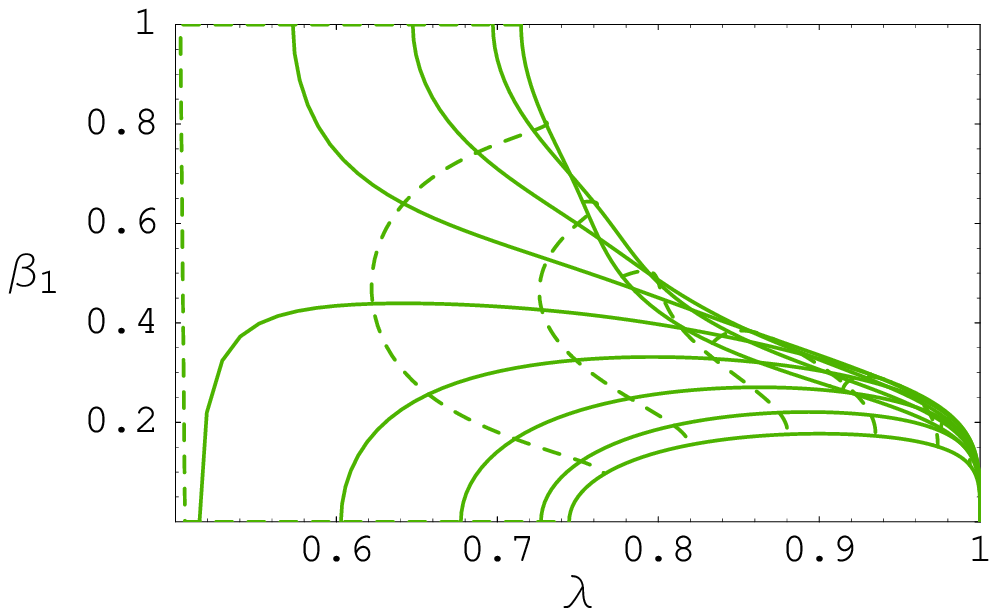}}
\caption{
 The creatable regions of the receiver's state space in the plane $(\lambda,\beta_1)$  for the chain of
$N = 2 N_1 = 40$ nodes: (a) the initial state $IS_1$; (b) the initial state $IS_2$;
(c) the initial state $IS_3$.}
 \label{Fig:reg3} 
\end{figure*}

The creatable regions corresponding to  $IS_1$,  $IS_2$ and  $IS_3$ are shown, respectively, in 
Fig.\ref{Fig:reg3}a,  Fig.\ref{Fig:reg3}b and 
Fig.\ref{Fig:reg3}c. This figure demonstrates the strong effect of the controlling state on the position and area of the creatable
region. On this figure,  the solid and dashed lines are the lines of constant, respectively, $\alpha_{12}$ and $\alpha_{11}$ which very  from 0 to 1 passing, respectively, 
(i) from right to left-downward  and
from left to right (Fig.\ref{Fig:reg3}a); therewith the line 
$\alpha_{11}=0$ covers the lines $\lambda=0$ and (partially) $\beta_1=0$, $\beta_1=1$,  and
(ii) from top to  bottom and from left to right (Fig.\ref{Fig:reg3}b,c)

\section{Symmetric communication line and simple quantum gates}
\label{Section:gates}
The remote creation of states may be considered a task in quantum
communication, as we did in the previous section. We now change
perspectives slightly and look at the symmetric communication lines of
Sec. \ref{Section:symmetric} from the point of view of quantum
computing, or quantum control. In that perspective, we assume the
senders $S_1$ and $S_2$ to initially contain some kind of input, while
the state of $R$ at a certain time $t_0$ is the output of a
(generalized) quantum gate. In what follows, $S_1$ is assumed to
contain the input data while $S_2$ is a generalized control. (Note
that in standard quantum algorithms the control qubit is usually
assumed to be in either of the two basis states $|0\rangle$ or
$|1\rangle$, while here $S_2$ has a vastly wider range of
possibilities.) 

The systems we 
 consider are the communication line in Fig.\ref{Fig:2channel}b with
 $N_1=20$ and the minimal system of Eq. (\ref{5sites}) which we will
 henceforth denote by $N_1=2$ for brevity, and three types of initial states:
\begin{eqnarray}
\label{IStr}
IS_{10}: &&\; {\mbox{ $|\Psi_{S_10}\rangle$ is state (\ref{inS}), (\ref{parameters}) and $|\Psi_{S_20}\rangle =|0\rangle$}}\\\nonumber
IS_{01}: &&\; {\mbox{ $|\Psi_{S_20}\rangle$ is state (\ref{inR}), (\ref{parameters}) and $|\Psi_{S_10}\rangle =|0\rangle$}}\\\nonumber
IS_{11}: &&\; {\mbox{ $|\Psi_{S_10}\rangle$ is state (\ref{inS}), (\ref{parameters}) and $|\Psi_{S_20}\rangle $ is state (\ref{inR}), (\ref{parameters})}}.
\end{eqnarray}
In $IS_{10}$ and $IS_{01}$ we explore the effect of exchanging input
and control, with one being a superposition state containing at most
one excited qubit and the other one being the ground state. $IS_{11}$
describes a situation where input and control states are identical
superpositions with one or two excitations. To characterize the
parameters $\beta_1$ and $\lambda$ of the output state (\ref{RhoR}) as
they depend on the nature of the input state we use the following notation:
\begin{eqnarray}
\beta_{ij}=\beta_1|_{IS_{ij}},\;\;\lambda_{ij}=\lambda|_{IS_{ij}}.
\end{eqnarray}
Throughout this section we restrict ourselves to positive real values of the
coefficients (\ref{inpar}), that is, $\phi_{ij}=0$ for all $i,j$.

\subsection{First gate: switching unitary transformations of receiver}
\label{Section:FirstGate}
\begin{figure*}
\subfloat[]{\includegraphics[scale=0.8,angle=0]{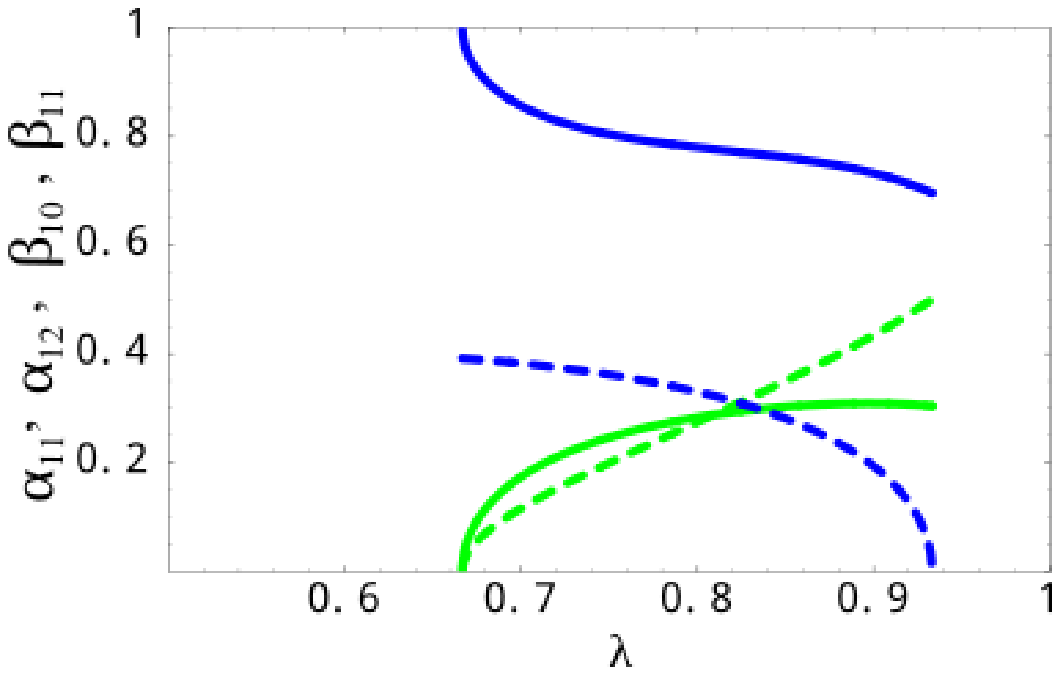}}
\subfloat[]{\includegraphics[scale=0.8,angle=0]{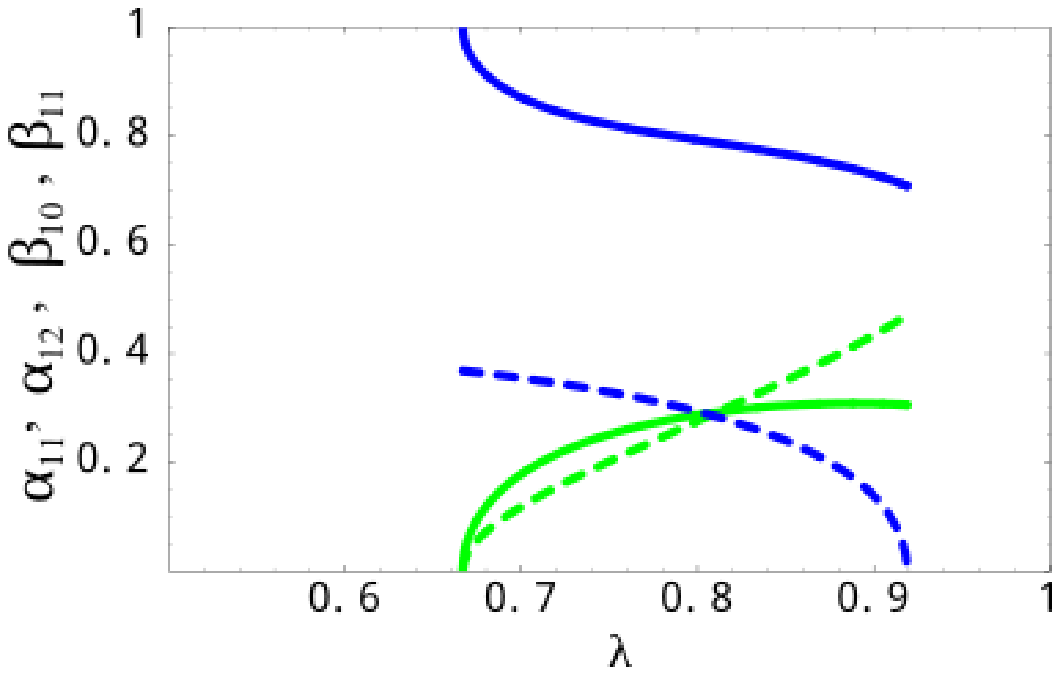}}
\caption{Parameters $\beta_{10}$ (solid lower line),  $\beta_{11}$ (solid upper line),
$\alpha_{11}$ (dashed increasing line) and $\alpha_{12}$ (dashed
decreasing line), for (a) the communication line of $N=2 N_1 +1=5$
nodes, and
(b) the communication line of $N=2 N_1 +1 =41$ nodes.
}
 \label{Fig:Gate1} 
\end{figure*}

We first assume that the nontrivial initial sender state is the same
in all three initial states (\ref{IStr}). That means, we have 
\begin{eqnarray}\label{ISpar}
\alpha_{2i}=\alpha_{1i}
\end{eqnarray}
between $|\Psi_{S_{10}}\rangle$ and  $|\Psi_{S_{20}}\rangle$ within
$IS_{11}$ and between   $|\Psi_{S_{10}}\rangle$ in $IS_{10}$ and
$|\Psi_{S_{20}}\rangle$ in $IS_{01}$. The output of the gate is the
mixed state $\rho^R$ (\ref{RhoR}) of the receiver, characterized by
parameters $\lambda$, $\beta_1$, and $\beta_2$. We concentrate on the
parameters  $\lambda$ and $\beta_1$, disregarding the phase angle
$\beta_2$. The action of the gate that we can achieve is described by
the mapping
\begin{eqnarray}\label{G1F}
&&IS_{10} \;\Rightarrow \; \rho^R(\lambda,\beta_{10}(\lambda))\\\nonumber
&&IS_{01} \;\Rightarrow \; \rho^R(\lambda,\beta_{10}(\lambda))\\\nonumber
&&IS_{11} \;\Rightarrow \; \rho^R(\lambda,\beta_{11}(\lambda)).
\end{eqnarray}
Thus, the eigenvalues of $\rho^R$ are the same in all three cases
($\lambda_{10}=\lambda_{01}=\lambda_{11}=\lambda$),  
while the parameter $\beta_1$ takes either of two values  $\beta_{10}$
or $\beta_{11}$, depending on the initial state. 
These values are uniquely related to $\lambda$, as  shown in Fig.\ref{Fig:Gate1}.
Thus, the considered gate switches between two possible sets of
eigenvectors of $\rho^R$ while keeping 
the same eigenvalues $\lambda$ and $1-\lambda$.

In order for the gate (\ref{G1F}) to work properly, the input
parameters $\alpha_{11}$ and $\alpha_{12}$ must have specific values
related to $\lambda$. 
Fig. \ref{Fig:Gate1} shows these values, along with the output parameters
 $\beta_{10}$, $\beta_{11}$   as 
functions of $\lambda$ for the chains of $N=5$ and $N=41$ nodes.

These figures show that there is a restricted interval  of the
parameter $\lambda$ where the gate  (\ref{G1F}) is realizable. In addition,
this interval  
is uniquely related to the  appropriate intervals of all other parameters: 
\begin{eqnarray}
\begin{array}{ll}
N=5&N=41\cr
0.666667 <\lambda    < 0.933064  \;\;\;&0.66667  <\lambda    < 0.918869 \cr
0 <\beta_{10} <0.304033\;\;\;&0 <\beta_{10} <0.305756 \cr
1>\beta_{11}>0.695726\;\;\;&1>\beta_{11} >0.708911\cr
0<\alpha_{11}<0.500114\;\;\;&0<\alpha_{11}<0.470208\cr
0.391827>\alpha_{12}>0\;\;\;&0.368291>\alpha_{12}>0.
\end{array}
\end{eqnarray}
Comparison of the intervals for the communication lines of $N=5$ and
$N=41$ nodes shows that there is a minor shrinking of all intervals
with an increase in $N$.

\subsection{Second gate: switching of eigenvector-parameter $\beta_1$ and eigenvalue $\lambda$}
\label{Section:SecondGate}
The second gate we study involves only two of the initial states
(\ref{IStr}) and its action on the output state is somewhat unusual:
\begin{eqnarray}\label{G2F}
&&IS_{10} \;\Rightarrow \; \rho^R(\lambda_{10},\beta_{10})\\\nonumber
&&IS_{11} \;\Rightarrow \; \rho^R(1-\beta_{10},\lambda_{10}).
\end{eqnarray}
In other words,
\begin{eqnarray}\label{G21}
&&\beta_{11} =\lambda_{10},\\\nonumber
&&\lambda_{11} = 1-\beta_{10}.
\end{eqnarray}
Thus, the  eigenvalue of the receiver's state corresponding to the
initial  state $IS_{11}$ is linearly expressed in terms of the  
eigenvector parameter $\beta_{10}$ of the receiver's state
corresponding to the initial  state $IS_{10}$, while the  
eigenvector-parameter $\beta_{11}$ is proportional to the eigenvalue
$\lambda_{10}$. Thus, the considered gate  
 switches parameters between eigenvalues and eigenvectors. The
 receiver's state corresponding to the  
 initial state $IS_{01}$ is not of  interest here, unlike with the
 first gate in Sec.\ref{Section:FirstGate}.

\begin{figure*}
\subfloat[]{\includegraphics[scale=0.8,angle=0]{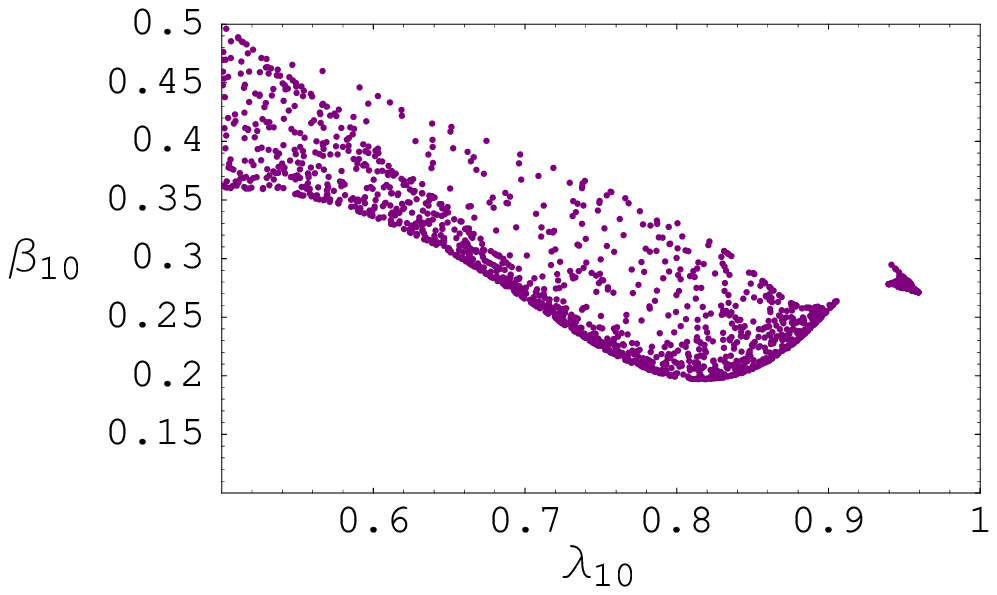}}
\subfloat[]{\includegraphics[scale=0.8,angle=0]{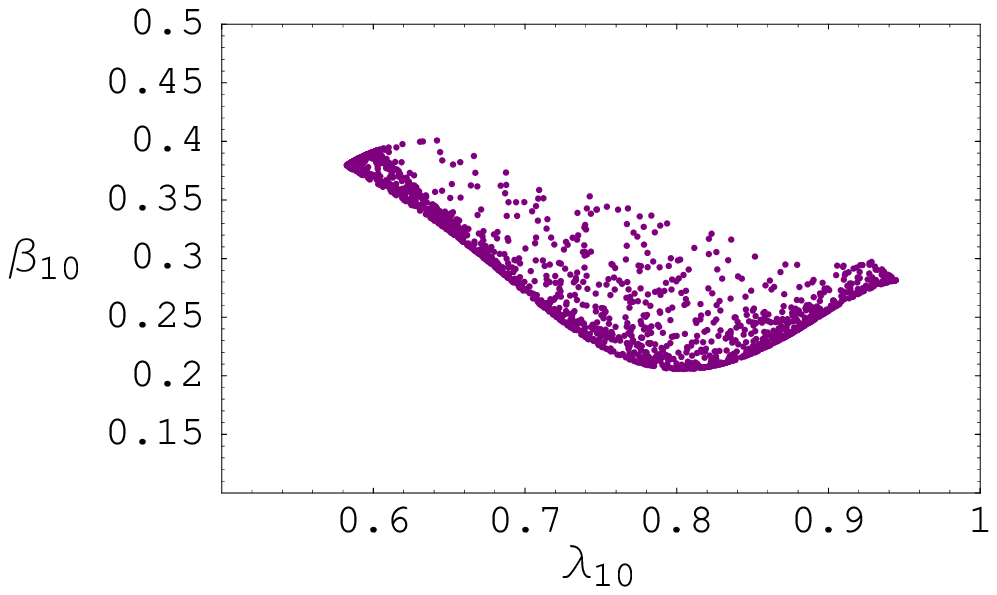}}
\caption{The domain of the function (\ref{G21}) in the plane
  $(\lambda_{10},\beta_{10})$, for
 (a) the communication line of $N=2 N_1 +1=5$ nodes, and
(b) the communication line of $N=2 N_1 +1 =41$ nodes.
}
 \label{Fig:LBN5N41} 
\end{figure*}

\begin{figure*}
\subfloat[]{\includegraphics[scale=0.8,angle=0]{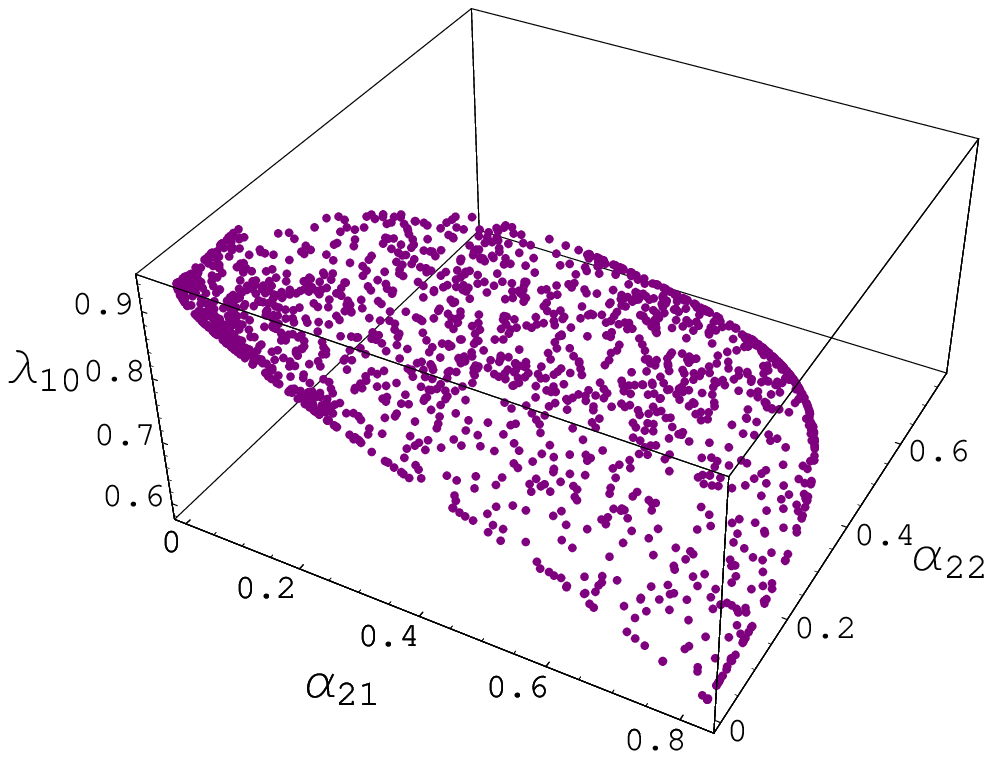}}
\subfloat[]{\includegraphics[scale=0.8,angle=0]{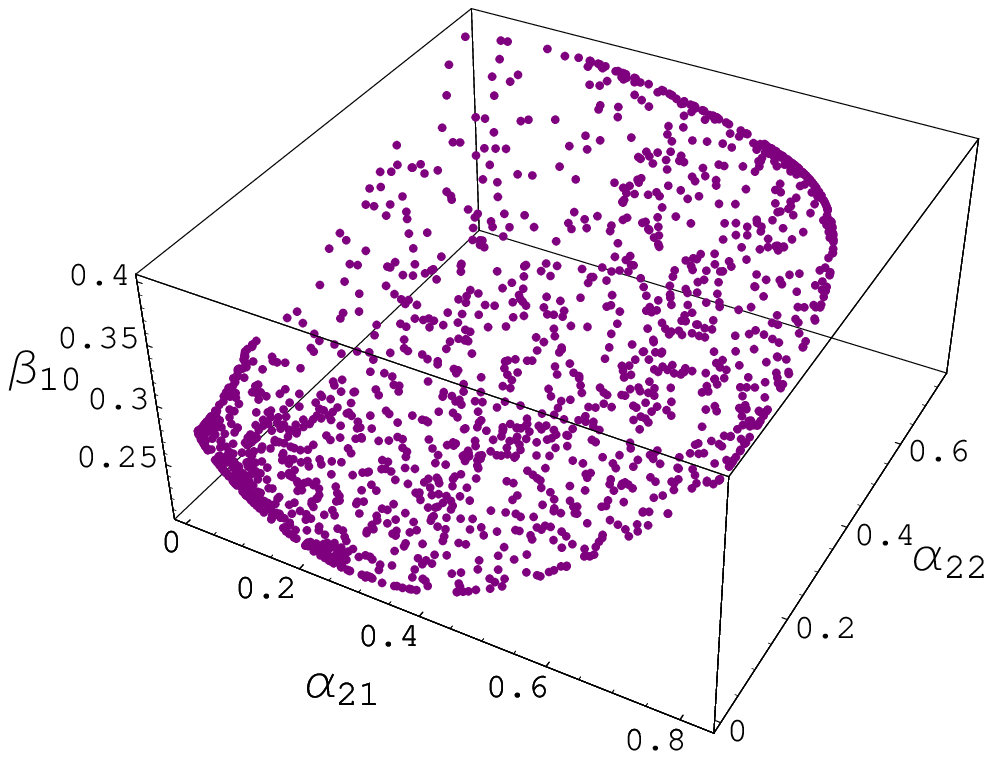}}\\
\subfloat[]{\includegraphics[scale=0.8,angle=0]{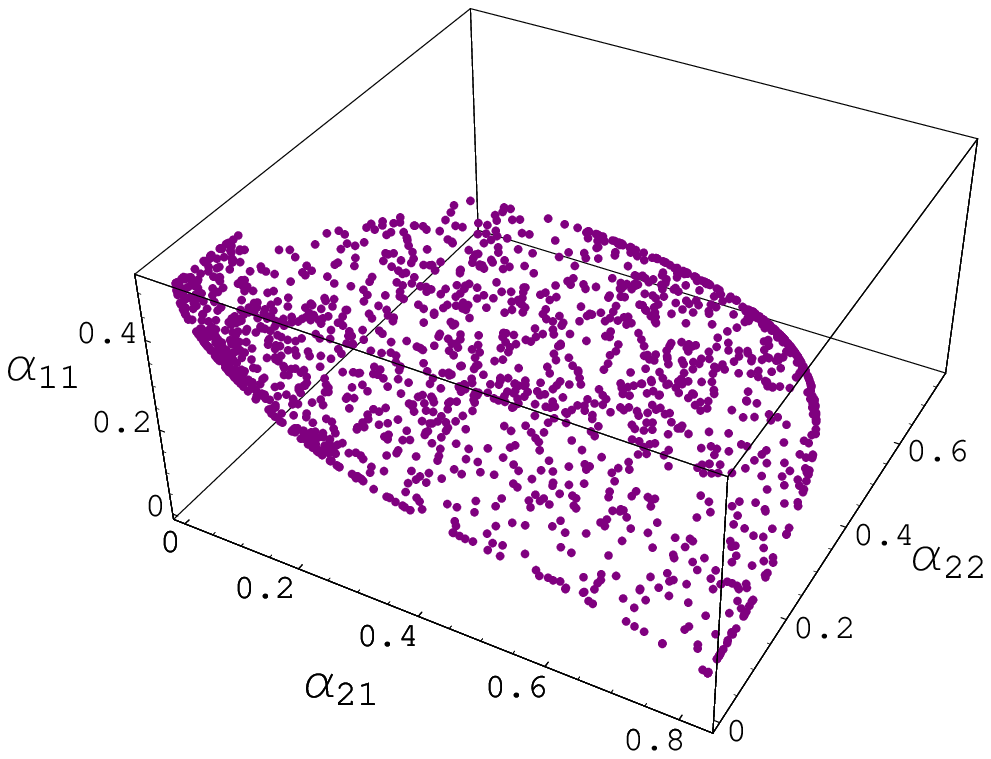}}
\subfloat[]{\includegraphics[scale=0.8,angle=0]{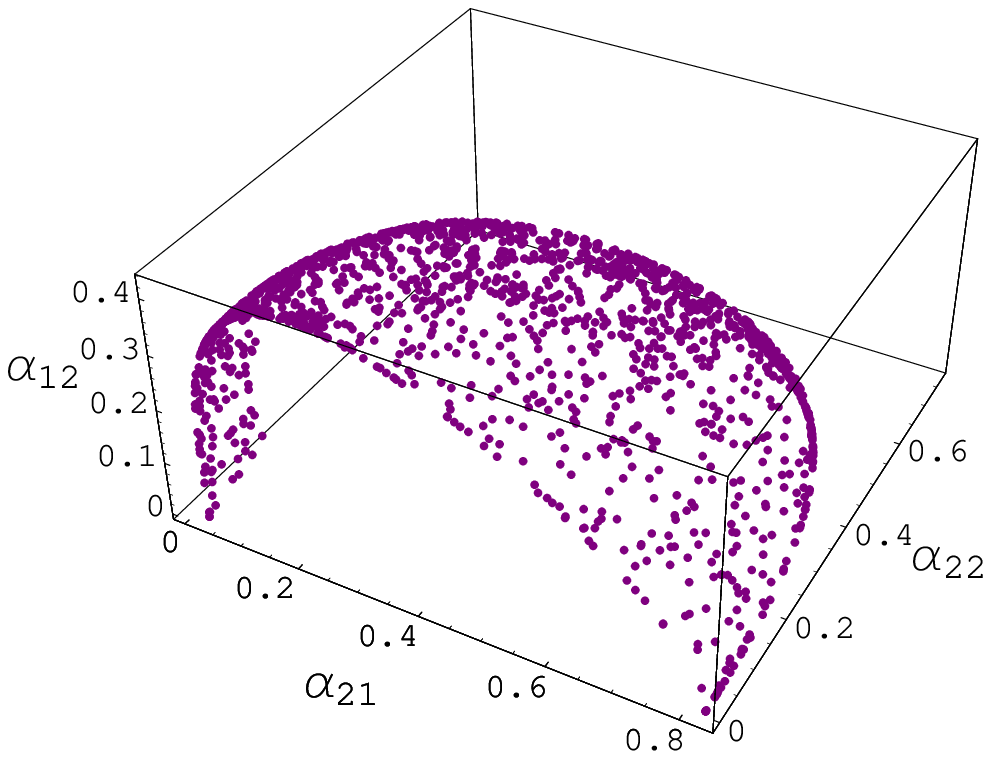}}
\caption{The   one-to-one
    {(except for the near-boundary region)} mapping (\ref{map}) 
for the communication line of $N=41$ nodes. (a) The parameter
$\lambda_{10}$, (b) the parameter $\beta_{10}$, (c) the parameter
$\alpha_{11}$, and (d) the parameter $\alpha_{12}$  over the plane 
$(\alpha_{21},\alpha_{22})$.
}
 \label{Fig:ALPN41} 
\end{figure*}

It is to be expected that the transformation (\ref{G21}) cannot be
achieved for arbitrary values of the input parameters. 
The domain of the function (\ref{G21}) in the plane $(\lambda_{10},\beta_{10})$ 
is shown in Fig.\ref{Fig:LBN5N41} for the communication lines of $N=5$
and $N=41$ nodes. 
These domains have been constructed using  random solutions  of system
(\ref{G21}). For this purpose, we consider the mapping (\ref{G21}) 
as a system of equations for the control parameters $\alpha_{ij}$
($i,j=1,2$; note that we no longer assume the relation (\ref{ISpar}) to
hold) and find 1567 (for $N=5$) and 1653 (for $N=41$) random
solutions. 
Each of the domains displayed in Fig.\ref{Fig:LBN5N41} 
  is creatable by the  parameters $\alpha_{ij}$ from  corresponding
  domains in the planes  
    $(\alpha_{11},\alpha_{12})$ and  $(\alpha_{21},\alpha_{22})$ and
    we observe that there are almost one-to-one {(except for the near-boundary subregion)} mappings  
\begin{eqnarray}\label{map}
&&
\{\alpha_{21},\alpha_{22}\} \;\;\Rightarrow \;\; \{\lambda_{10},\beta_{10}\}\\\nonumber
&&
\{\alpha_{21},\alpha_{22}\} \;\;\Rightarrow \;\; \{\alpha_{11},\alpha_{12}\}.
\end{eqnarray}
These mappings  are represented in  Fig.\ref{Fig:ALPN41} for the case $N=41$, where the parameters  $\lambda_{10}$,
$\beta_{10}$, $\alpha_{11}$ and $\alpha_{12}$ are shown as  surfaces over the plane of the parameters
$\alpha_{21}$ and $\alpha_{22}$. 
 The discontinuity of the domain in Fig.\ref{Fig:LBN5N41}a at $\lambda_{10}\sim 0.92$ is related to the form of 
the creatable subregion of the receiver's state space.
 
\section{Conclusion}
\label{Section:conclusion}

We have considered the problem of remote state creation using two
senders connected to a single receiver by boundary-optimized channels. 
Such a communication line can be used, for instance, to share  the
creatable region of the receiver's state space between two senders so
that the subregions creatable by each of the sender (provided that the
other one is in the ground state) do not overlap. Therefore, if the
state is registered at the receiver, we know from which sender  the
creation process originated. Varying the phase of the initial state we
can change the position of the creatable subregions.   

Another application of the two-sender communication line is a quantum
gate operating on  the mixed state of the receiver. In that case, the
initial state of one of the senders is regarded as an input signal,
the initial state of the other sender is a controlling signal. The
state of the receiver at the proper time instant is the output
signal. There is a large family of functions which can be realized by
such a gate of which we here consider only two functions. 

In the first function the controlling state changes  the parameter of the eigenvector matrix of the receiver state keeping the eigenvalues unchanged.
In the second function the controlling state switches parameters
between the eigenvectors and eigenvalues of the output state. 
Each of these functions can be  realized if 
the control parameters $\alpha_{ij}$ of the initial state and the 
creatable parameters of the receiver's state belong  to the proper domain of their values;  we characterize all these domains. 
Further study of such gates would be  of interest.

This work is partially supported by the program of RAS ''Element base of quantum
computers'' and by the Russian Foundation for Basic Research, grant No.15-07-07928.

\end{document}